\documentclass[conference]{IEEEtran}
\IEEEoverridecommandlockouts

\usepackage{amsmath,amssymb,amsfonts}
\usepackage{algorithmic}
\usepackage{graphicx}
\usepackage{biblatex}
\usepackage{textcomp}
\usepackage{xcolor}
\def\BibTeX{{\rm B\kern-.05em{\sc i\kern-.025em b}\kern-.08em
    T\kern-.1667em\lower.7ex\hbox{E}\kern-.125emX}}
\newtheorem{definition}{Definition.}
\newtheorem{theorem}{Theorem.}

\begin{document}

\title{RPoA: Redefined Proof of Activity}

\makeatletter
\newcommand{\linebreakand}{%
  \end{@IEEEauthorhalign}
  \hfill\mbox{}\par
  \mbox{}\hfill\begin{@IEEEauthorhalign}
}
\makeatother

\author{
\IEEEauthorblockN{Sina Kamali}
\IEEEauthorblockA{\textit{School of ECE} \\
\textit{University of Tehran}\\
Tehran, Iran \\
kamali.sina@ut.ac.ir}
\and
\IEEEauthorblockN{Shayan Shabihi}
\IEEEauthorblockA{\textit{School of ECE} \\
\textit{University of Tehran}\\
Tehran, Iran \\
shabihi@ut.ac.ir}
\and
\IEEEauthorblockN{Mohammad Taha Fakharian}
\IEEEauthorblockA{\textit{School of ECE} \\
\textit{University of Tehran}\\
Tehran, Iran \\
fakharian.taha@ut.ac.ir}
\and
\IEEEauthorblockN{Alireza Arbabi}
\IEEEauthorblockA{\textit{School of ECE} \\
\textit{University of Tehran}\\
Tehran, Iran \\
alireza.arbabi@ut.ac.ir}
\linebreakand
\IEEEauthorblockN{Pouriya Tajmehrabi}
\IEEEauthorblockA{\textit{School of ECE} \\
\textit{University of Tehran}\\
Tehran, Iran \\
pouriyatajmehrabi@ut.ac.ir}
\and
\IEEEauthorblockN{Mohammad Saadati}
\IEEEauthorblockA{\textit{School of ECE} \\
\textit{University of Tehran}\\
Tehran, Iran \\
mohammadsaadati80@ut.ac.ir}
\and
\IEEEauthorblockN{Behnam Bahrak}
\IEEEauthorblockA{\textit{School of ECE} \\
\textit{University of Tehran}\\
Tehran, Iran \\
bahrak@ut.ac.ir}
}

\maketitle

\begin{abstract}
The consensus protocol is the core of a blockchain system which guarantees its secure and stable operation. Proof of Activity (PoA) is a consensus protocol that tries to address some of the issues pertinent to the most widely used protocols, such as Proof of Stake (PoS) and Proof of Work (PoW). However, it still needs to solve the issues regarding high energy consumption, significant resources required, high mining latency, and the need for private blockchains.

In this paper, we propose Redefined Proof of Activity (RPoA), a new consensus protocol that builds on top of some of the best features of the existing protocols, such as PoW, PoS, and PoA, and values active service provided by users on the network. Our approach tries to address the issues above and falls in the service-based protocols category that gives mining credit to users as they serve on the network.
\end{abstract}


\section{Introduction}

The consensus protocol is the core feature of a cryptocurrency. Thus, defining it is one of the most critical tasks in designing a new cryptocurrency because most essential features of a blockchain network, such as decentralization, security, and throughput, are tied closely to the consensus protocol. Hence, a suitable consensus protocol can guarantee blockchain systems' fault tolerance and security.

Launched in 2009, Bitcoin \parencite[]{nakamoto2008bitcoin} continues to proliferate. At its core, Bitcoin uses \textit{Proof of Work} (PoW) \parencite[]{perez2019analysis} as its consensus protocol. Bitcoin is inflation resistant, which is a direct result of block rewards decaying over time with respect to the block height \parencite[]{derks2018chaining}, thus making the total amount of Bitcoins available finite. Other impressive features of Bitcoin include decentralization, operational security against most prominent attack vectors, and integration with other on-chain and off-chain protocols. Regarding security, Bitcoin is resistant against $>50\%$ adversarial power commonly known as the infamous 51\% attack \parencite[]{chohan2021double} \parencite[]{foaobstudy}. 

Recently \textit{Proof of Stake} (PoS) protocols have emerged as an energy-efficient alternative to PoW \parencite[]{zhang2020evaluation} \parencite[]{deb2021posat}. Ethereum \parencite[]{wood2014ethereum} used to run on PoW but has recently transitioned to PoS \parencite[]{kapengut2022event} for its lower power consumption, better throughput, and more flexibility \parencite[]{yang2019delegated}. Ethereum introduces \textit{smart contracts}, a technology upon which all transaction-based state machines could be developed. Smart contracts let users develop decentralized applications, also known as DAPPs and decentralized autonomous organizations (DAOs) \parencite[]{antonopoulos2018mastering}. Such features let developers also build alternative blockchain-based systems on Ethereum's smart contracts. Over the years, PoS has been proven to be susceptible to many forms of alternative attacks, including Nothing-at-Stake attack \parencite[]{houy2014will}, Reorg and Liveness attacks \parencite[]{schwarz2021three}, and Avalanche attack \parencite[]{neu2022two}.

The staking system is a significant part of the PoS consensus protocol. Staking is used to freeze assets on the network. One use case of staking in the PoS systems is giving decision-making power to entities with a stake in the system. The term \emph{validator} refers to the mentioned entities in most PoS networks. Validators collect transactions from a shared pool, cumulate them into new blocks, and broadcast them to the network. 

Mainstream protocols such as PoW or PoS still have drawbacks \parencite[] {oyinloye2021blockchain} \parencite[] {shifferaw2021limitations}. Apart from energy consumption, PoW \parencite[]{ante2021bitcoin} suffers from low throughput \parencite[] {bez2019scalability}. Nothing-at-Stake is a major venerability of PoS, where the same stake can be used to grind on the many blocks. \textit{Proof of Activity} (PoA) \parencite[] {bentov2014proof} is another consensus protocol that combines PoW and PoS into a more robust and secure mechanism. In PoA, the mining process begins similarly to PoW, with various miners trying to outpace each other based on their computation power to find a new block. When a new block is mined, the system switches to PoS, where a group of validators is pseudo-randomly selected to validate the block. Although PoA provides more security and decentralization than its discussed counterparts, it still suffers from issues such as outrageous energy consumption\parencite[]{zhang2020evaluation} and a high mining latency \parencite[]{schwarz2022stochastic}.

In this study, we introduce Redefined Proof of Activity (RPoA), a novel consensus protocol that combines the best features of PoW, PoS, and PoA. RPoA gives additional mining power to users based on their accumulated activity. We define \textit{activity} as a factor that measures the amount of service a user provides to the network. RPoA provides most of the features mentioned above while adding extra features such as reduced energy consumption, lowered hardware requirements for entry-level mining, and the same level of security as its predecessors. Furthermore, RPoA encourages users to stay active on the network, thus keeping it alive, which propels users not to leave and stay active and make more contributions to the network.

\textit{Our contributions}: we propose a consensus protocol that defines \textit{activity} as a measure of users' dedication to the network and, in comparison with the existing solutions, provides the same level of security with much lower power consumption, better supports dynamic availability, features less mining latency, correctly supports non-private blockchains, and requires less computational resources from new miners.


As a consensus protocol, RPoA could be suitably used within any compatible decentralized framework, whether they are based on smart contracts, blockchains, or block trees. Furthermore, many different technologies, including zk-SNARKs \parencite[]{bitansky2012extractable} and cryptographic accumulators \parencite[]{ozcelik2021overview} could be integrated with RPoA for better performance and improved security.

The rest of the paper is structured as follows: Section II reviews related studies. Section III introduces RPoA, discusses its features, and explains the mining equation. Section IV investigates RPoA security and how previous security issues are addressed. Section V further discusses the computation of upload fees and how transaction fees are used. Section VI discusses possible future directions and concludes the paper.

\section{Related Work}

Not much research exists on activity and valuable work being considered the main consensus of a decentralized network. One of the major protocols that came out with a similar idea is proof of useful work (PoUW) \parencite[]{lihu2020proof}, which encourages miners to perform useful work, such as training machine learning models in exchange for a reward. For example, PoUW may require users to pay fees for submitting their models to the network and bring miners to train such models later and serve as \emph{useful entities} on the network. Additionally, \textit{proof of prestige} (PoP) \parencite[]{krol2021proof} has been introduced that values unverifiable tasks in the network. PoP introduces a new concept called \textit{prestige}, a volatile resource that regenerates over time. Prestige is gained by performing \textit{useful work} and is spent while deriving benefits from the provided services. Other similar protocols include consensus protocols in \textit{File-Coin} \parencite[]{benisi2020blockchain}, \textit{Golem} \parencite[]{golem}, decentralized mining pools like \textit{SmartPool} \parencite[]{luu2017smartpool}, and networks that reward entities bridging on-chain and off-chain protocols.

RPoA seeks to address issues of similar protocols and attempts to amend some of their major characteristics to perform better in many ways. In contrast to many other protocols, RPoA's \textit{activity} function could be defined in several ways as applicable to each application domain and thus provides better compatibility with the currently existing chains. For example \parencite[]{lihu2020proof}\parencite[]{benisi2020blockchain}\parencite[]{luu2017smartpool} only define domain-specific \textit{value} functions. In addition, compared to PoA and PoW, our proposed approach offers much lower energy consumption. Therefore, it is more scalable and environment-friendly due to the net effect of \textit{activity} in lowering the mining difficulty. Furthermore, our work offers more simplicity and is easier to implement compared to PoA and PoP, which incorporate more complicated and costly algorithms \parencite[]{bentov2014proof}\parencite[]{krol2021proof}. RPoA also supports integration with most external ecosystems such as \textit{Decentralized Finance} \parencite[]{chen2020blockchain} and other previously discussed protocols, a concept that makes RPoA flexibly applicable to be used in real-world networks.

While we address some significant drawbacks of most current systems, like disregard for users' valuable work and vast energy inefficiency, we also do not devote ourselves to other principal features a consensus protocol must support. Our design additionally focuses on much easier implementation, fairer rewards by using Geometric Rewarding \parencite[]{fanti2019compounding}, better integration with currently available on-chain and off-chain networks, and incentivizing users to better contribute to the network over time. It also displays resistance against most prominent attack vectors, one of the main features every mainstream consensus protocol should have.

\section{Main Network Mechanism}
The following subsections will discuss the base subsystems for RPoA to run, including the staking, activity, and mining subsystems.

\subsection{Staking System}
Staking freezes assets for a set period to help bankroll the network. RPoA requires users to stake service fees as refundable assets on the network, which keeps the network funded. It also indirectly incentivizes users to stay more active, which helps further grow the network for an increased worth of their assets when unfrozen. In other words, there are no non-refundable service fees in the RPoA; every paid service fee is refundable and released after a certain period. 

The target network's characteristics decide how staked assets are recorded on the corresponding blockchain. For example, UTXO-based networks could assign certain transactions for staking the required fees and have nodes verify that all fees are already staked before recording \textit{service transactions}. In contrast, networks based on smart contracts could use contracts between the two parties to help ensure the essential fees are paid in advance. Other systems could also verify staked fees similarly and thus be compatible with RPoA.

\subsection{Activity System}

RPoA prioritizes users based on the amount of their activity on the network. Doing so incentivizes users to stay active on the network. Thus, to track users' activities, at least one specific transaction type with measurable worth should be defined. This type of transaction will be referred to as \textit{service transaction} in the rest of this paper. These types of transactions should be defined network-specifically, some examples of which include \textit{file-upload transactions} in file-sharing networks and \textit{gamble transactions} in gambling networks.

As previously discussed, service transactions assign activities to users. Therefore, these transactions must have some essential characteristics. First, a \textit{worth} function should be defined, measuring each transaction's relative worth in the system. In RPoA, the \textit{activity} function computes this relative worth value per service transaction at any particular time. Second, there must be a way to track such transactions and validate every prior service transaction a user has made. To do so, RPoA records service transactions on the blockchain to be later verified by the active nodes.

The \textit{activity} function computes the relative weight of every service transaction. We first define two helper functions, $TF$ and $WF$, that respectively weigh the effects of elapsed time and relative worth of each service transaction, and later define \textit{activity} as computed combining these factors.

\begin{definition}
Let $p$ be the period passed from the creation of the service transaction, and let $T, r \in  \mathbb{R}^+$ be arbitrary constants that change how the time factor decays with respect to time. With this, time factor $TF(p)$ is defined as follows:
\end{definition}

\begin{equation}
TF(p) = (\frac{T}{p+T})^r
\end{equation}

\begin{definition}
Let $\chi, L \in  \mathbb{R}^+$ be two arbitrary constants. $WF(w)$ is the worth factor of a representative service transaction worth $w$ and is defined as follows:
\end{definition}

\begin{equation}
WF(w) = (\frac{-L}{w+\frac{L}{\chi}})+\chi
\end{equation}

\begin{figure*}[htbp]
    \centerline{\includegraphics[width=15cm]{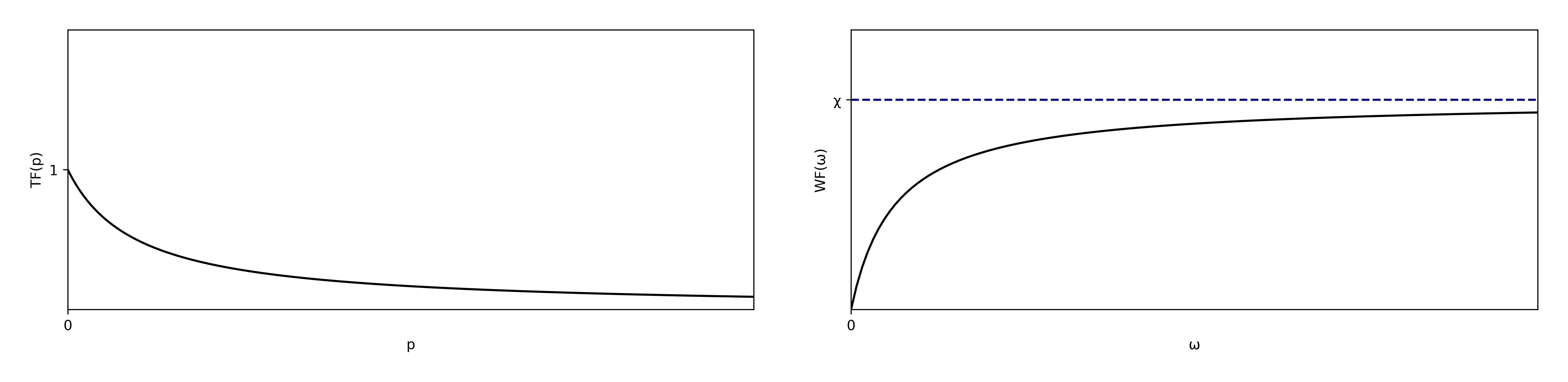}}
    \caption{The distribution of WF and TF with respect to corresponding parameters}
    \label{fig:WFTF}
\end{figure*}

\begin{definition}
Following the definitions for $TF$ and $WF$, we denote service transaction activity by $\xi(p, w)$, which is defined as follows:
\end{definition}

\begin{equation}
\xi(p, w)= \alpha \times TF(p).WF(w)
\end{equation}
While with $p\rightarrow \infty$, $TF$ approaches a value of 0, with $w\rightarrow \infty$ the $WF$ approaches $\chi$. Thus, the cooperative effect of the two functions multiplied at $p \rightarrow 0$ and $w\rightarrow \infty$ would impose a max assigned activity of $\alpha.\chi$ per transaction.

Following the provided formulation for activity assigned to each service transaction, we compute the total activity per user by taking a summation or exponential average over their entire activity history. Here we opt for the former, but the latter option could also be considered depending on the characteristics of the target network.

\begin{definition}
Let $S_u$ be the set of all service transactions of a user $u$, and $p, w$ be the same as used above. The total activity of $u$ has a base value of 1 and is denoted by $\psi_u$, which is defined as follows:
\end{definition}

\begin{equation}
\psi_u=1 + \sum_{i\in S_u}{\xi(p_i, w_i)}
\end{equation}

\subsection{Mining System}

\textit{Mining} is the process in which a miner proposes new blocks whose hashes satisfy a criterion. More precisely, a block, in this sense, is generated via the accumulation of a series of transactions taken from a common pool. Miners are rewarded in two ways for mining each block: There is a block reward in place, and on top of that, users can put extra tips called \textit{miner wages} in transactions similar to Bitcoin to encourage miners to prioritize them.

RPoA uses a \textit{cryptographic puzzle} requiring new block hashes to be under a network-determined value. To maintain a constant rate for new block generations, the network keeps a factor called the \textit{mining difficulty} leveled. This factor controls the complexity of the mining puzzle so that new blocks are statistically added at determined intervals, called $CT$ hereafter,  and based on the existent hash rate. The following will mathematically define the mining formula and the difficulty factor.

\begin{definition}
A function $h$ takes a single byte array input and generates the $\beta$-bit hash of it as output.
\end{definition}

The mining formula represents the hashing puzzle and must put linearly-growing chances of mining for miners with higher activities. It should also toughen the puzzle, given the difficulty factor increases and vice versa.
\begin{definition}
Let $BH$ denote the block header hash to be found. Also, let $t$ and $u$ denote the given time and mining user. The following defines the mining inequality:
\end{definition}

\begin{equation}
    h(BH) \leq 2^{\beta - 1} \times \eta(t) . \psi_u
\end{equation}
In the above inequality, $2^{\beta-1}$ is the initial difficulty the network starts with, providing a low starter difficulty for network initialization. Also, in the same context, $\eta$ and $\psi$  are \textit{difficulty factor} and \textit{activity} functions of inputs $t$ and $u$, respectively. The following provides the formal definition for \textit{difficulty factor}:
\begin{definition}
Let $\upsilon(t)$ be the average time it takes to mine a block in the network at time $t$. $\eta(t)$ is the difficulty factor at time $t$ and is defined as follows:
\end{definition}

\begin{equation}
    \eta(t) = \frac{\upsilon(t)}{CT}
\end{equation}

\begin{theorem}
Let $\zeta(t)$ be the number of total network transactions, and $n_u$, $n_b$, and $n_m$ denote the number of network users, blocks, and miners, respectively. Also, Let $\Omega(t)$ denote the total network cumulative activity at time $t$, and $\Gamma$ denote the constant block size. Considering $\Xi_u$, the maximum level of activity bound to each user, and $\zeta_u(t)$, the number of previous transactions per user, is a normal distribution with mean $\zeta(t)/n_u$, $\Xi_u$ is directly dependant upon $n_b$, and inversely upon $n_m$, and is independent of the parameters used in calculation of $\psi_u$. Further proof is provided in Appendix 1.
\end{theorem}

\section{Security}
One of the significant features of a consensus protocol is resiliency against challenging attack vectors, and RPoA is no exception. In the current section, we elaborate on the qualities that make RPoA a trustworthy and secure alternative.

Like any other currently available major protocol, RPoA is vulnerable to $>50\%$ attacks \parencite[]{foaobstudy}. However, the power of a node cannot be as defined as the hashing power or total stake they have, like in PoW and PoS, respectively. To thoroughly discuss the $>50\%$ attack, we first give the following definition for a user's \textit{power}. 

\begin{definition}
    Let $g(u)$ denote the hashing power of a user and $U$ be the set of all users. $\Pi(u, t)$ is the power of a user $u \in U$ at a given time $t$ and is defined as follows:
\end{definition}
\begin{equation}
    \Pi(u, t) = \frac{g(u).\psi(u)}{\sum_{u' \in U}{} g(u').w(u')}
\end{equation}
With the power of a user-defined as above, one can now observe that with an adversary party with over half of the network's total power, $\sum_{u \in U}{}\Pi(u, t)$ can have a critical dishonest impact on the network, most potentially leading to double spending and other related attacks \parencite[]{chohan2021double}

The nothing-at-stake phenomenon is a well-known issue, majorly targeting PoS systems \parencite[]{li2017securing} \parencite[]{nguyen2019proof}. Unlike in PoW, in PoS, it is not computationally costly for validators to add new blocks to the blockchain, a problem commonly referred to as the nothing-at-stake security issue. The issue theoretically arises anytime a fork in the blockchain, either due to a malicious action or by accident, when more than one validator simultaneously proposes a new, valid block. The RPoA's mining system is fundamentally different from that of PoS. It is much closer to PoW in that miners generate hashes suiting the criteria imposed by the \textit{difficulty factor} at the time of mining. Hence, the nothing-at-stake attack does not fundamentally apply to the case of RPoA. Similarly, other pertinent attacks such as the \textit{Avalanche Attack} \parencite[]{neu2022two} are not relevant to our case too.

The Sybil attack \parencite[]{douceur2002sybil}, yet another important and well-studied attack on consensus protocols, allows users to gain additional power by simply using a single node to operate several active fake identities or so-called, \textit{Sybil identities}. It is also an issue specific to non-PoW-based systems; With PoW in effect, one has to delegate their computational power to the multiple Sybil identities, resulting in no increased hashing power for the whole mining party, giving them no additional control in the network. To overcome the potential challenges imposed by a Sybil attack practiced on the network, RPoA charges new users a time-variable fee for their entry to the network. This fee is charged upon making the first service transaction and is called the \textit{entrance fee}.

\section{Upload Fees}
By charging fees for service transactions, RPoA provides resistance against not only Sybil attacks but also constraints on the amount of assigned activity to the contributors. First, entrance fees play a vital role in preventing Sybil attacks and, when optimized for the target network, can prevent such attacks entirely. Definition 9 provides a formulation for entrance fees. Furthermore, dynamic service transaction fees are charged per service transaction. RPoA charges higher transaction fees the higher the activity of a user, thus limiting the levels of activity a user can reach. While most protocols distribute transaction fees between the contributing miners, RPoA has users stake fees on the network until a certain network-specific time. Lastly, users can provide extra fees in their service transactions to encourage miners to mine their transactions faster. This kind of fee is called the \textit{miner wage} and is implemented at a network-specific level.

\begin{definition}
Let $H(t)$ denote the latest block height at entrance time $t$, and $\gamma$ be the base entrance fee for users' attendance. The entrance fee $E(t)$ is defined as follows:
\end{definition}

\begin{equation}
    E(t) = \gamma \times \sqrt{H(t)}
\end{equation}
By using this entrance mechanism, we prevent adversaries from evading the activity factor of the service transaction fee by incorporating new identities. Keep in mind that by utilizing the block height of the time of registration, entering the network becomes more challenging as time progresses, which incentivizes the users to keep the same identity and thus its activity and stakes for as long as possible. 

\begin{definition}
Let $\Omega$ denote the maximum service transaction worth allowed per block, and $\alpha$ be the base fee for a service transaction. The base service transaction fee $\beta(w)$ is the base transaction fee of a transaction with a specific amount of worth $w$:
\end{definition}

\begin{equation}
\beta(w) = \alpha \times \frac{w}{\Omega}
\end{equation}

\begin{definition}
Let $\gamma$ be an arbitrarily large constant base fee, $w$ be the transaction worth as noted above, and $u$ denote the specific user in concern for the upload fee calculation. With $\psi(u)$ denoting the activity of the specified user at time t, $F(u, w)$ gives the net payable upload fee and is defined as:
\end{definition}

\begin{equation}
    F(u, w) = \gamma \times \beta(w) \times \psi_u
\end{equation}
As stated earlier, transaction fees help maintain a balance between the activity level a user gains by pumping up fees as users' activity grows. In other words, the more activity one achieves, the more fee one has to pay for their activity cumulations and spend more. This increase in fees prevents users from making transactions back-to-back.

\section{Conclusion and Future Work}
In this paper, we introduced the RPoA consensus protocol, which incentivizes users to stay active on the network while not discouraging new users from joining it and tries to solve the major issues of previous protocols. RPoA merely needs a decentralized medium to run on and therefore supports most current systems such as smart contract- and UTXO-based ones. For future developments, we foresee RPoA applied to decentralized finance systems, and majorly to smart contract-based ones, such as in state-of-the-art flash loan \parencite[]{wang2020towards} and liquidation systems \parencite[]{jensen2021leveraged}. Furthermore, the concept of service transaction fees contributing to the network's staked assets significantly reduces the network's non-liquidity. The staked capital could also be processed for revenue generation, which could support the network in many ways. Similar ideas have been previously implemented, such as in REX coin \parencite[]{rex}. We strongly believe that the RPoA could be useful in many situations, including the ones above.

\section*{Acknowledgment}

We express gratitude to the contributors who helped significantly in this project: Sahar Shirmardi and Shamim Nasiri for their kind support in the formulation of \textit{activity}, and Ali Ebrahimi for assistance in the development of a proof of concept system based on RPoA.


\printbibliography

\appendix
\section{Proofs}
\subsection{Proof of Theorem 1}
In this proof, let $\nu={\frac{n_b}{n_u}}$ and $\Gamma$ be the constant block size of the network. If $\zeta_u(t)$ is a normal distribution with mean $\nu$, following equation (11), we know that the expected value of $\zeta$ grows as a linear function of $\nu$. 
\begin{equation}
E(\zeta_u(t)) = \frac{E(\zeta(t))}{n_u}  = \frac{E(c\times n_b.\Gamma)}{n_u}   = c' \frac{n_b}{n_u}
\end{equation}
Considering $\alpha.\chi$ equals the max activity gained by making a single service transaction for a user, and $\Xi_u$ as the maximum bound activity level per user, we have:

\begin{equation}
E(\Xi_u) = E(\zeta_u.\alpha.\chi)
\end{equation}
From equation (12), assuming that $\alpha$ and $\chi$ are predefined, network-specific constants, one can conclude that the $E(\Xi)$ is also a linear function of $\nu$, and therefore the theorem is proved. In this sense, the $E(\Xi)$ is irrelevant to the parameters used in the definition of $\psi_u$.

\end{document}